\documentclass[prl,amsmath,amssymb,reprint,nofootinbib]{revtex4-1}

\usepackage{graphicx}
\usepackage{color}
\usepackage{amsmath, slashed}

\usepackage{graphicx}
\usepackage{epsfig}
\usepackage{amsmath}
\usepackage{amsfonts}
\usepackage{amssymb}

\usepackage{braket}
\usepackage{cancel}
\usepackage{pstricks}
\usepackage{color}
\usepackage{feynmf}

\usepackage[normalem]{ulem}

\def\lsim{\:\raisebox{-0.5ex}{$\stackrel{\textstyle<}{\sim}$}\:}

\def \n{\noindent}
\def \BR{\cal{B}}
\usepackage{epstopdf}

\begin{document}

\title{Resonance enhancement of Charm CP}
\author{Amarjit Soni}
\address{$^1$Physics Department, Brookhaven National Laboratory, Upton, NY 11973, USA}

\begin{abstract}

It is suggested that a nearby $0^{++}$ resonance, $f_0$(1710) of mass 
$m_f=1723 MeV$ and width $\Gamma=139$ MeV is playing a significant role in efficiently providing the strong (CP-conserving) and weak (CP-odd) phase simultaneously in the recently observed direct CP asymmetry $\Delta A_{CP}$ by the LHCb collaboration. The direct CP arises by the well known penguin-tree interference wherein the virtual b-quark in the c-u penguin is the source of the Kobayashi-Maskawa CP-odd  phase, $\gamma$, in the SM. Loop (penguin) corrections generate  left-right operators enhancing coupling to the $0^{++}$ scalar resonance. The scalar resonance is likely rich in gluonic content perhaps leading to a better understanding of large breaking of flavor SU(3) that has been known for a long-time.
Approximate calculations give a rough understanding of the observed size of the CP asymmetry. The mechanism leads to  several interesting implications which can be experimentally studied and tested. 
Moreover, in an analogous fashion to $f_0$, 4-quark operators also generate $P \times P$, P being a pseudo-scalar bilinear, which may be dominated by the nearby $\eta (1760)$ of width about 250 MeV that can influence final states such as 4 $\pi$'s, $\eta^{\prime}(\eta) + \pi^+ + \pi^-$ etc which could exhibit CP violating triple correlation or energy asymmetries. We also briefly discuss CP violation in radiative charm decays and suggest that simple final states
$\gamma \rho$ and $\gamma \phi$ are best suited for sizeable asymmetries as well as providing precise tests of the SM.

\end{abstract}

\maketitle

\section{Introduction \& motivation}

Recently LHCb announced the exciting discovery of direct CP violation in charm decays~\cite{LHCb_CP2019},
\begin{align}
\Delta A_{CP} \equiv [A_{CP}(K^+K^-)-A_{CP}(\pi^+\pi^-)] \nonumber \\
=(-15.4 \pm 2.9)X10^{-4}
\label{DeltaACP}
\end{align}
giving a 5.3 $\sigma$ signal of {\it direct} CP violation in charm decays. There are at least two reasons why it is very important to seek a clear and quantitative understanding of the origin of this asymmetry. For one thing,  we do not understand how baryogenesis arises as the SM-CKM~\cite{NC63,KM72} phase falls short by many orders of magnitude in accounting for this phenomena. Moreover, naturalness arguments strongly suggest that generic type of new physics beyond the SM should entail new CP-odd phase(s). The importance of the  naturalness reasoning that suggests that new physics to entail new CP-odd phase can hardly be overemphasized. In this context, it is useful to remember the analogy with neutrino mass. In the 70's and 80's neutrino oscillation searches used to give null results because experiments in those days were not sensitive enough to probe light enough $\delta m^2$ region. But we kept on pushing the experimental frontiers and  improving the limits as there was no good reason for the neutrino mass to be zero.
Finally that long and arduous search paid off in the 90's. The rationale for the quest for BSM-CP should be stronger as we have already known for a long time that CP is not a good symmetry of nature~\cite{BNL_1964}. We simply ought to continue vigorously to improve our bounds and our theoretical understanding.
Unfortunately accurate quantitative estimates of direct CP asymmetries are a very difficult challenge as they invariably entail non-perturbative QCD dynamics.

In the case of charm decays, it is worth noting (as will be briefly explained below) that if the hints of new physics in semi-leptonic B decays,
$B \to D \tau (l=e, \mu) \nu$ are confirmed, then that new physics could entail BSM-CP-odd phase(s) that can intervene in the processes that are our focus.

Specifically, for $D^0 \to K^+ K^-$ and $\to \pi^+ \pi^-$ it appears that a novel role is  being played by a neighboring resonance, $f_0(1710)$~\cite{cPDG2018} carrying quantum numbers of the vacuum, in communicating the essential CP-odd (weak) and CP-even (strong) phase that are mandatory to drive the CP asymmetry. Fortunately,
the mechanism readily renders several testable predictions though
large data samples will be needed. Given the LHC, LHCb run plans and possible upgrades along with recent commissioning of the SUPER-KEK-B factory and the Belle-II detector with the anticipated significant increase in luminosity, timing is rather ripe for further investigations that may enrichen our understanding.

\section{Estimates}
As is well known, in charm decays, direct CP asymmetries can only arise through the interference of the tree decay
amplitudes with the penguin (loop) amplitude~\cite{BSS_79}. In the SM especially in charm decays the loop is essential as it actually allows the three generations to contribute as is needed due to the Kobayashi-Maskawa requirement~\cite{KM72} through the mechanics of the CKM-mixing matrix~\cite{AJB01}; see Fig 1. For convenience, it is best to absorb the part of the penguin contribution with
the virtual light s and d quarks into the tree amplitude. Thus, for Cabibbo suppressed final states that are of interest here,

\begin{figure}[hbt]
\begin{centering}
\includegraphics[width=0.5\textwidth]{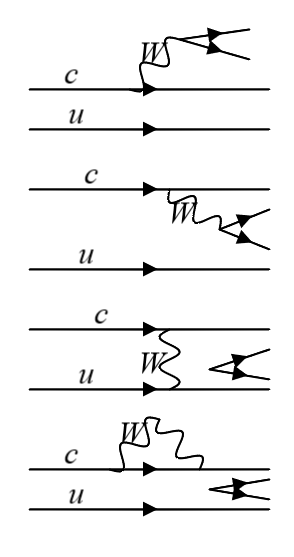}
\par\end{centering}
\caption{\label{fig:Feyn4}
        Some diagrams contributing to charm decay; gluons causing pair creation are not shown. Top to bottom: color-allowed tree; color-suppressed tree; weak annihilation and penguin. Color allowed tree is relevant to modes such as $K^+ K^-$ and $\pi^+ \pi^-$, color-suppressed one for $\pi^0 \pi^0$ and weak-annihilation for $K_s K_s$. Since $\alpha_{s(m_{charm})}$ is not that small, these distinct topologies especially for the color-suppressed tree and the weak annihilation can receive large corrections due to final state interactions.
    }
\end{figure}
    
\begin{equation}
A_{hh} = \lambda T_{hh} + \lambda^5 P_{hh}e^{i[\delta_{st}+\eta_{wk}]}
\end{equation}
where $\lambda=0.2245$ is the Cabibbo~\cite{NC63} angle in the Wolfenstein~\cite{Wolf84} representation, $T_{hh}$ is the tree amplitude arising from the matrix elements of the usual 4-quark left $left X left$ ($LXL$) operators, $Q_1$ and $Q_2$, and $P_{hh}$ is the penguin amplitude arising from the QCD-penguin operators $Q_3$ to $Q_6$; for definiteness, we follow the notation of~\cite{ASW80} but see also~\cite{Golden:1989qx} and ~\cite{BLB_RMP}. Electroweak penguins are ignored throughout as their contributions in charm decays is extremely small.
Amongst the penguin operators, $Q_5$ and $Q_6$ are $LXR$ and  under Fierz rearrangment give rise to scalar and pseudo-scalar operators~\cite{SVZ76}. In particular,  the scalar will couple to the $0^{++}$ resonance, f(1710) mentioned before, following the
weak decay. Its width provides a good way to encapsulate the final state interaction (CP-conserving) phase~\cite{EHS91,EHS92,AS94}, $\delta_{st}$. The penguin also has a CP-odd phase as there is a $V_{ub}$ coupling
involved which is the source of the unitarity angle $\gamma \approx 73.5$ degrees ~\cite{cPDG2018} and in here $ \sin \eta_{wk} \approx \sin \gamma \approx 0.96$ and in the resonance ($f_0$) dominance approximation, $ \sin \delta_{st}$ is estimated to be $\approx 0.7$.
The partial rate asymmetry defined as usual as (with $\BR$~ being the branching ratio~$\equiv$~Br),
\begin{align}
\alpha_{PRA} =  \BR [I \to F ] - \BR [\bar I \to \bar F]/ \BR [I \to F ] + \BR [\bar I \to \bar F]
\nonumber\\
\end{align}
is then given for $D^0 \to h^+ h^-$  by,
\begin{eqnarray}
\alpha_{hh} = 2 \lambda^4 P_{hh} \sin(\delta_{st}) \sin (\gamma)/T_{hh}
\label{PRA}
\end{eqnarray}

 \n In charm decays (unlike K or B), it is exceedingly difficult for the penguin correction  to make a sizable contribution to the total rate given the intrinsic 
CKM-suppression factor of $\lambda^4$. In fact it is worth noting that even for K-decays,  lattice simulations ~\cite{RBC-UKQCD-2015,RBC2001} have clearly demonstrated that once one uses a renormalization point above around 1.5 GeV  the penguin contribution to
the rate for $K \to \pi \pi$, in the isospin $ = 0$ channel, is very small 
$\lsim 1 \% $.

Thus, in charm decays, one can safely assume tree contribution dominates. This has the immediate consequence that the PRA~\cite{AS_CKM2018},

\begin{eqnarray}
\alpha_{hh} \propto 1/\sqrt{\BR}
\end{eqnarray}

\n It will be explained a bit later that this simple expectation gets modified 
appreciably because the resonance is influencing  the numerator of (\ref{PRA}).

It is to be stressed again that reliable method(s) for calculating matrix elements for such purely hadronic final states do not exist and at best we can make some rough estimates. Retaining only the factorizable matrix elements leads to an approximate expression for the tree amplitude,

\begin{eqnarray}
t_{hh} = (G_F/2\sqrt(2))(C_2 + C_1/3) f_K f(0) m^2_D
\end{eqnarray}

\n where $G_F$ is the fermi constant,  C's are the Wilson coefficients, $f_K$ is the kaon pseudoscalar decay constant,
f(0) is the $D \to K$ semi-leptonic form factor at $q^2 = 0$~\cite{FLAG2019} and $m_D$ the $D^0$ mass. For numerical purposes a renormalization point, $\mu =2 GeV$
is used throughout.

In the resonance approximation, the corresponding   (real part) of the penguin amplitude is roughly estimated as,

\begin{align}
P_{hh} = (G_F/2\sqrt(2)) C_6 A^2 \sqrt(\rho^2 + \eta^2) \nonumber \\
(f_D m_D/m_c) K_f m_f^2
\end{align}

\n where $\eta$, $\rho$ and A are the usual Wolfenstein parameters, $f_D$ the decay constant~\cite{FLAG2019}, $m_c$ the charm quark mass
(at $\mu$) and $m_f$ the mass of the $f_0$ and $K_f$ is $\approx 3.7$ estimated from the width of the $f_0$.

Using all the above input into (\ref{PRA}) one finds,
\begin{eqnarray}
\alpha_{K^+ K^-} \approx 5.5 \times 10^{-4}
\end{eqnarray}

\n In deducing this we have used from~\cite{cPDG2018} $\Gamma(f_0 \to KK)/\Gamma_{f_0} \approx 0.4$.

\subsection {Interplay of CPT  \& resonant CP}

Because of the CPT constraint that life-time of particle must equal to that
of its antiparticle, we must have, 

\begin{eqnarray}
\sum_X \Delta\Gamma(X)=0
\label{pra:general}
\end{eqnarray}

\n  where X are the various final states that emerge from the decays of $D^0$ and

$\Delta\Gamma(X) =\Gamma(D^0\to X)- \Gamma(\overline 
D^0\to \overline X)$

Since at the quark level in charm decays there are only two channels, $c \to u \bar d d$ and $c \to u \bar s s$, this means because of CPT we must have~\cite{AS_PTEP13,HG87}, 
\begin{eqnarray}
\Delta \Gamma (c\to u \bar d d) = - \Delta \Gamma (c \to u \bar s s)
\end{eqnarray}
\n  At the meson level, it is suggested~\cite{AS_PTEP13} that this materializes into,
\begin{eqnarray}
\Delta \Gamma (\pi^+ \pi^-) = - \Delta \Gamma ( K^ + K^-)
\end{eqnarray}

\n Using the tree dominance, as emphasized above, one then arrives at the relation,
\begin{eqnarray}
\alpha_{K^+ K^-} \propto - \alpha_{\pi^+ \pi^-}/\sqrt{2.8}
\end{eqnarray}
where, we have taken for simplicity that the two Brs differ by about $\sqrt{2.8}$~\cite{AS_CKM2018}.

However, f-dominance of the penguin amplitude, in the h h channel that is our central focus, appearing in the numerator of (\ref{PRA}) modifies this expectation appreciably. This is because, it appears that ~\cite{cPDG2018}.
\begin{eqnarray}
Br(f_0 \to \pi \pi)/Br(f_0 \to KK) \approx 0.4
\end{eqnarray}

\n Thus, since the penguin amplitude entering the numerator of (\ref{PRA})
is dominated by the $f_0$ and its denominator tends to go as $\sqrt(Br)$, we find,
\begin{eqnarray}
- \alpha_{\pi^+ \pi^-} / \alpha_{K^+ K^-} = \sqrt(0.4     \times 2.8) \approx 1.06
\end{eqnarray}

\n though, it must be stressed that the central value of 0.4~\cite{cPDG2018} has large uncertainty. The minus sign is based on expectations from CPT~\cite{AS_PTEP13}. This has the important consequence that $\Delta A_{CP}$ (\ref{DeltaACP}), as defined by LHCb,
is actually larger in magnitude than $\alpha_{K^+K^-}$ or $\alpha_{\pi^+\pi^-}$.
For LHCb their  $\Delta A_{CP}$  is an observable that allows them to overcome the problem that LHC is a pp machine (so the initial state is not self-conjugate) and related background issues. 
A determination of the individual asymmetries, $\alpha_{K^+K^-}$ and $\alpha_{\pi^+ \pi^-}$, should be given  high priority by LHCb as well as by Belle-II; some experimental efforts in this direction are already underway~\cite{JB2017,LHCb2017, Belle2017}.

\section{implications and possible tests}

The resonance dominance idea has several interesting experimental consequences.

\begin{itemize}
    \item $f_0$ is likely rather rich in gluonic content~\cite{MP99}.
    This may also facilitate an understanding of the $f_0$ branching ratio being larger into $K^+ K^-$ than into $\pi^+ \pi^-$ since
    low energy QCD dynamics suggests that the couplings of scalar and pseudo-scalar glueballs to quarks (and mesons) is likely mass dependent ~\cite{CS84,hyc2010}.  Moreover, it may also at least partly explain why the $D^0$ branching ratio (which as explained in
    preceding paras is dominated by the tree operators, $Q_1$ and $Q_2$) exhibits
    such a large breaking of SU(3).  Factorization does suggest $Br(D^0 \to K^+ K^-)/Br(D^0 \to \pi^+ \pi^-) \approx [(f_K/f_{\pi}) \times (f_0(m_K^2)/f_0(m_{\pi^2})]^2 \approx 1.7$
    which still falls short of the 2.8 seen by experiment. We suggest that this long-standing difficulty has its origin in our resonance hypothesis. The point is that while it is readily seen that $0^{++}$ operators such as $Q_6$ cause mixing with $f_0$ that does not mean that the tree operators $Q_1$, $Q_2$ will not mix with scalars. Perturbation theory suggests a (QCD) loop-suppressed mixing for the $L \times L$ tree operators; however,  this is likely an under-estimate given the rich gluonic content of the $f_0$.

    \item Over the years, the PRA for $D^0 \to K_s K_s$, $\alpha_{K_s K_s}$,  has been of considerable interest~\cite{AS_PTEP13,GLR2013, NS_17} as it is expected to be larger than in $K^+ K^-$. With the resonance idea isospin dictates that 
    the corresponding penguin amplitude appearing in the numerator of
    (\ref{PRA}) will be smaller by a factor of two. On the other hand, the
    branching ratio into 2 $K_s$ is smaller than into $K^+ K^-$ by a factor of about 23~\cite{cPDG2018}. Thus we should expect,
    \begin{align}
    \alpha_{K_s K_s}/\alpha_{K^+ K^-} \approx \sqrt(23/4) \nonumber \\
    \approx 2.4
    \end{align}
    
    \item In this resonance hypothesis one can also find a relation between the PRA for $D^0$ to 2 $\pi^0$ versus to $\pi^+ \pi^-$. Due to isospin we should expect amplitude for $f_0$ to two $\pi^0$ to be approximately the same as to charged pions.
    Given the tree dominance in $D^0$ decays and using the known branching ratios into neutral and charged pions
    (this ratio is about 0.56), one can deduce that
    $\alpha_{\pi^0 \pi^0} = 1.3~ \alpha_{\pi^+ \pi^-}$. 
    
    \item In an analogous fashion to $f_0(1710)$, the 4-quark operators  also generate terms that go as  $P \times P$ such as  $ (\bar u \gamma_5 c) (\bar u \gamma_5 u)$ which lead to a $D^0$ transition to $\eta(1760)$, possibly contributing dominantly to final states of the type $\eta^{\prime}(\eta) \pi^+ \pi^-$,
    2 $\pi^+$ 2 $\pi^-$ etc. These could exhibit other types of CP asymmetries such as energy or triple correlation but to demonstrate that will, of course, require flavor tagging of the initial $D^0$.

\end{itemize}

\begin{figure}[htb]
    \begin{centering}
    \includegraphics[width=0.5\textwidth]{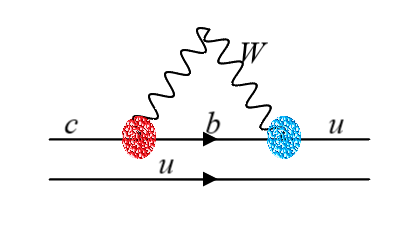}
    \par\end{centering}
    \caption{\label{fig:BpenguinRB}
        b-penguin in charm decay; left c-b (red) vertex may have a CP-odd phase endowed by new physics affecting $R_{D(*)}$ whereas the right b-u (blue) vertex contains the SM-CKM phase.  
}
\end{figure}

\section{$R_D(*)$ anomaly and new physics in b-penguin contribution to charm decays~\cite{AS_CKM2018}}

In charm decays there are as always three standard penguin contributing amplitudes.
They proceed via a virtual d-quark,  a virtual s-quark and a virtual b-quark. In the SM with the CKM matrix in the Wolfenstein convention the b-penguin is the one that carries the dominant CP-odd phase at the $V_{ub}$ vertex (see Fig~\ref{fig:BpenguinRB}); in fact to a good approximation this is the angle $\gamma$ of the unitarity triangle, as already identified in the discussion above. It is interesting to note that in the past $\approx$ 5 years there has been considerable excitement as experiments at Babar~\cite{BabarRD,BABARRD2}, Belle~\cite{BelleRD,BelleRD2,BelleRD3,BelleRD4} and LHCb~\cite{LHCbRD,LHCbRD2} involving the studies of simple semi-leptonic B-decays via $B \to D^{*} l (\tau) \nu$ have been exhibiting signs of lepton universality violations (LUV) amounting to about 3 $\sigma$. If these signs of possible new physics withstand further scrutiny and do become a reality then because of naturalness arguments as alluded to before there may be an important impact on the b-penguin in charm decays as then the c-b vertex therein may well carry a BSM-CP-odd phase as well; see Fig~\ref{fig:BpenguinRB}. Regrettably because of our inability to calculate    things reliably,  it may not be easy  to quantitatively discern the effects of a new CP-odd phase from that of the CKM-phase; however, it is hoped that lattice and/or phenomenological techniques can address some of the relevant issues. (See below)

\section{ Brief remarks on radiative charm decays}

CP violation in radiative decays of charm mesons has been of considerable interest in recent years~\cite{IK2012, LZ2012, AS_PTEP13, BH2017, BH2018, AHT2018}. We limit our discussion to a few remarks. First is to draw attention to a new class of radiative decays of D's to heavy nearby resonances. Two good examples are 
$\phi(1680)$ and $\rho(1700)$~\cite{cPDG2018}. The $\phi$ has a mass of about 1680 MeV and width around 150 MeV and predominantly decays to $K K^*(892)$ and the $\rho$ has a mass of 1702 MeV, width around 250 MeV and decays to $\rho \pi \pi$, 2 ($\pi^+ \pi^-$) etc. These radiative decays are  accompanied by rather soft photons with maximum energies of about 300 MeV. The branching ratios of these modes should also be around a few $\times 10^{-5}$ roughly similar to $\gamma \rho(770)$ and $\gamma \phi(1020)$~\cite{cPDG2018} and likely somewhat bigger. The dominant production mechanism is weak annihilation tree graph
(See Fig 1 as for 2 $K_s$)
and receive sub-dominant contributions from radiative penguins. Just as in the case of 2 $Ks$ these should exhibit sizeable CP asymmetries,
perhaps O(few $\times 10^{-3}$). Depending on the final state they can lead to energy and triple correlation 
asymmetries as well. 
It is important to note that the PRA in $\gamma \rho$ and $\gamma \phi$ will have opposite signs because of CPT, as explained above. It is likely therefore that there  are advantages in defining, as in $\Delta A_{CP}$, difference in radiative asymmetries,
\begin{align}
\Delta A^{\gamma}_{CP} =
\alpha_{\gamma \phi} - \alpha_{\gamma \rho}
\end{align}
It is also likely that CP asymmetry, $\alpha_{\gamma \rho}$ will be larger than $\alpha_{\gamma \phi}$ since the Br of  the former is smaller and also because the $\bar u u$ pair in the penguin needs to convert to $\bar s s$ for the case of $\phi$. Lastly, an important challenge for theory is to render these into precision test of the SM. We turn to this issue briefly below.

\section{Possible role for the lattice}

Lattice techniques are currently being used~\cite{RBC-UKQCD-2015,Ishizuka2018} to address $K \to \pi \pi$ and the direct CP violation parameter, $\epsilon^{\prime}$, for that. These are based on the Lellouch-Luscher~\cite{LL01} method which cannot be used as such for D decays as many multi-particle states lie below $m_D$;  nevertheless efforts are continuing~\cite{HS2012,BH2015} to address this difficulty. Be that as it may, 
it  may sill be  possible to study the weak transition of $D^0 \to f_0(1710)$ or $D^0 \to \eta(1760)$
using lattice techniques though this may have its own challenges. At this point in time, known lattice methods ~\cite{BHS1993,PB2014,JLAB2016,JLAB2017,CK2018,SM2018} appear best suited to provide the SM expectation for $\alpha_{\gamma \rho}$ and  $\alpha_{\gamma \phi}$ wherein phenomenological approaches may also have a good chance for some success~\cite{ABS1994,KSW1997,BFS2001,KM10}.

\section{Summary}
Summarizing, it is suggested that resonances near $D^0$ may be facilitating the partial rate asymmetry measured recently by LHCb. The scalar state $f_0(1710)$ carrying quantum numbers of the vacuum is likely dominating decays to $K^+ K^-$ and $\pi^+ \pi^-$ while the neighboring pseudo-scalar state $\eta(1760)$ may be dominating multi-particle (4 pions) final state. As such the $\eta (1760)$ is likely entering the dynamics of other asymmetries such as energy or triple correlation in those final states. We also briefly discuss CP  violation in radiative charm decays and in particular emphasize that PRA's in $D^0 \to \gamma \rho$ and to $\gamma \phi$ should be sizable and seem most promising for providing precise tests of the Standard Model. Lastly, it is to be stressed that the $f_0(1710)$ and $\eta(1760)$ discussed here should be viewed as illustrations of the important effects nearby resonances can have on CP asymmetries. It has been known for a long-time that the charm region is rich with resonances and in fact there may be more that are not there yet~\cite{BES2005, LHCb2014} in~\cite{cPDG2018} but if confirmed, could have important impact on quantitative discussions on CP asymmetries in charm decays.

\section{Acknowledgements}
Its a pleasure to thank Sheldon Stone for many useful comments and suggestions and also to Angelo Di Canto and Tim Gershon for useful discussions and encouragement. This research was supported in part by the U.S. DOE contract 
\#DE-SC0012704.

\medskip
{\bf Note added I}

While this paper was being prepared~\cite{AL2019,PA2019,GS2019,LDY2019,ZZX2019} appeared also addressing~\cite{LHCb_CP2019}. \\

{\bf Note added II}

We briefly recall that few years ago there were some experimental hints for a somewhat larger $\Delta A_{CP}$~\cite{LHCb2011,CDF2012}. This led to intense theoretical activity~\cite{Bigi2011,Isidori2011,BBGR2012,FNS2012,Li2012,Brod2012,Grossman2012,hyc2012A,hyc2012B,AS_PTEP13,KP2017} resulting in a better understanding of the predictions of the SM.

\bibliographystyle{unsrtnat}
\bibliography{mybib2}

\end{document}